\documentclass[preprint,pre]{revtex4}
\usepackage[T1]{fontenc}
\usepackage[latin1]{inputenc}
\usepackage[dvips]{graphicx}
\usepackage{float}
\usepackage{amssymb}
\usepackage{amsmath}

\newcommand{\lav}{\left\langle}
\newcommand{\rav}{\right\rangle}

\restylefloat{figure}

\begin{document}
\title{Phase transition in the two-component symmetric 
exclusion process with open boundaries}
\author{A. Brzank\textsuperscript{1,2} and
G. M. Sch\"utz\textsuperscript{1,3}}
\affiliation{\textsuperscript{1}Institut f\"ur Festk\"orperforschung, Forschungszentrum J\"ulich, 52425 J\"ulich, Germany\\
\textsuperscript{2}Fakult\"at für Physik und Geowissenschaften, Universit\"at Leipzig,
Linnestrasse 5, D-04103 Leipzig, Germany\\
\textsuperscript{3}Interdisziplin\"ares Zentrum f\"ur Komplexe Systeme, Universit\"at Bonn,
53117 Bonn, Germany}
\email{brzank@rz.uni-leipzig.de, g.schuetz@fz-juelich.de}
\date{\today}

\begin{abstract}
We consider single-file diffusion in an open system with two species $A,B$ of 
particles. At the boundaries we assume different reservoir densities which 
drive the system into a non-equilibrium steady state. As a model we use an
one-dimensional two-component simple symmetric exclusion process with two 
different hopping rates $D_A,D_B$ and open boundaries. For investigating the 
dynamics in the hydrodynamic limit we derive a system of coupled non-linear 
diffusion equations for the coarse-grained particle densities. The relaxation 
of the initial density profile is analyzed by numerical integration. Exact 
analytical expressions are obtained for the self-diffusion 
coefficients, which turns out to be length-dependent, and for 
the stationary solution. In the steady state we find a discontinuous 
boundary-induced 
phase transition as the total exterior density gradient between 
the system boundaries is varied. At one boundary a boundary layer develops 
inside which the current flows against the local density gradient. Generically
the width of the boundary layer and the bulk density profiles do not depend on 
the two hopping rates. At the phase transition line, however, the individual
density profiles depend strongly on the ratio $D_A/D_B$. Dynamic Monte Carlo 
simulation confirm our theoretical predictions.\\[4mm]
{\it Keywords:} Driven diffusive systems (Theory), Exact results, Stochastic particle dynamics (Theory)
\end{abstract}
\maketitle

\section{Introduction}

Single-file diffusion occurs when a random motion of particles is
confined to a narrow channel where they cannot pass each other due to hard-core repulsion.
Examples for this behaviour include molecules diffusing in the channels of 
certain zeolites \cite{Karg92,Kukl96,Keil00}, transport of colloidal spheres 
in one-dimensional channels \cite{Wei00}, or the dynamics of ``defects'' 
inside the ``tube'' to which entangled polymers are confined \cite{Perk94}. 
Also biological motors moving along macromolecules or microtubuli exhibit 
single-file motion, see e.g. \cite{Schu97,Lipo05,Nish05} and 
references therein. Single-file diffusion is an effectively one-dimensional 
many-body random process which exhibits intriguing correlation effects. 
It is known for a long time that the mean-square displacement of a tagged 
particle grows subdiffusively for late times with the square root of time 
\cite{Levi73}, see also \cite{vanB83} for an insightful theoretical derivation. 
Also macroscopic 
stationary properties in open systems, driven by an external gradient of the 
chemical potential shows interesting long-range correlations even if the 
interaction between particles is strictly local \cite{Spoh83,Derr02,Bert02}.
This implies failure of Onsager theory and indicates strong non-equilibrium 
behaviour.

So far most theoretical treatments of single-diffusion have focussed on
indistinguishable particles. The reference model for this process is the 
one-dimensional symmetric exclusion processes (SEP), a lattice model where 
hard-core particles hop randomly to nearest-neighbour sites, provided the 
target is empty \cite{Ligg99,Schu01}. However, already an experiment aimed at 
verifying the subdiffusive behaviour of a single tracer particle requires to 
tag one or more particles to make them distinguishable. This is done by
labelling some particles without changing their diffusion properties and thus 
corresponds to a two-species SEP with identical hopping rates. A 
rigorous treatment of the macroscopic behaviour of such a 
two-component SEP with many tagged particles in terms of coupled non-linear 
diffusion equations for an infinite system was given by Quastel 
\cite{Quas92}. A finite system with open boundaries was treated only very 
recently \cite{Brza06}. 

More generally, however, two physically distinct species of particles 
may simultaneously move inside the same 
channel. In recent measurements a mixture of toluene and 
propane was adsorbed into different zeolites \cite{Czap02}. The authors 
measured the temperature dependent outflow and  noticed a trapping effect due 
to single-file diffusion: In zeolites with long intracrystalline channels
the stronger adsorbed toluene molecules control due to blocking the outflow of 
propane. In terms of the SEP the different physical properties of toluene and 
propane require two different hopping rates.
Also studying the drift velocity of an entangled polymer in gel 
electrophoresis in the framework of reptation theory necessitate a description 
of the defect dynamics in terms of two distinct species of particles 
\cite{Rubi87,Duke89}. 

Somewhat surprisingly the hydrodynamic theory of two-component lattice 
gas models seems quite undeveloped even though there is some
recent progress \cite{Toth03,Popk03b,Schu03}. In \cite{Keil00} 
Keil et al. review the Maxwell-Stefan theory describing the diffusive 
behaviour of a binary fluid mixture where the total current, i.e. the sum 
of both species, is zero. The particle-particle interaction is taken into 
account by including a friction between the species being proportional to 
the differences in the velocities. However, this approach is purely 
phenomenological and does not provide a link between microscopic
properties of the system and the transport coeffficients entering the
macroscopic equation. In particular, this approach is not adequate for 
single-file diffusion in a finite system where the self-diffusivity of 
single particles has turned out to play an important role \cite{Brza06}.

In this paper we go beyond Ref. \cite{Brza06} by considering a two-component
symmetric exclusion process with {\it different} hopping rates 
(defined in Sec. 2) and by studying not only stationary properties but also 
the time evolution of the particle density and relaxation towards the 
stationary state.  To this end, generalizing the arguments put forward in 
\cite{Quas92} for identical particles, we consider open boundaries where 
both ends of the chain are connected to particle reservoirs with different 
chemical potentials. We derive in Sec. 3 a set of coupled nonlinear 
diffusion equations and we analyze
the time-dependent solutions of these PDE's by numerical integration.
This is compared with simulation data obtained from dynamical Monte-Carlo
simulations (DMCS) of the process.  In Sec. 4 the stationary solution 
is obtained analytically in closed form and discussed in some detail.
Some conclusions are drawn in Sec. 5. The exact computation of the
self-diffusion coefficients for the two particle species is presented in
the first appendix. The second appendix provides for self-containedness
some technical details of that derivation.

\section{Two-component symmetric exclusion process}

\begin{figure}
\centerline{\includegraphics[width=14cm]{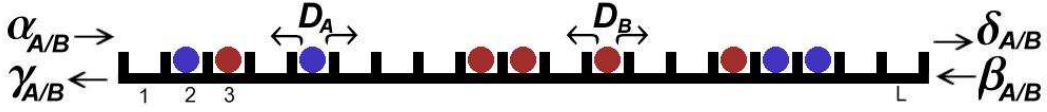}}
\caption{Two-component symmetric exclusion model with open boundaries.
Each species of particles has its own hopping rates $D_A,D_B$. At the
boundaries particles are extracted and injected with rates as indicated.}
\label{latticeThreeState}
\end{figure}

Following \cite{Brza06} we consider a one-dimensional lattice with $L$ lattice 
sites (Fig.~\ref{latticeThreeState}). Each site $i$ can be empty or occupied by a 
particle of type $A$ or $B$. Due to hard-core interaction any site carries at most 
one particle. Particles can hop to nearest neighbour sites (provided the target 
site is empty) with hopping rates $D_{A/B}$ which are the ``bare''
diffusion coefficients, i.e., the proportionally factor in the mean-square
displacement if no other particles were present. Jump events occur after an 
exponentially random time which in dynamic Monte Carlo simulations (DMCS) 
is modelled
by random sequential update \cite{Brza06}. 

Let $a_i$ ($b_i$) count the $A$ ($B$) particles on site $i$. Then the
densities are the ensemble-averaged expectation values of the 
respective particle occupation numbers:
$\lav a_i \rav \equiv \rho_A(i)$, $\lav b_i \rav \equiv \rho_B(i)$. 
The probability of finding no particle at site $i$ is 
$\lav v_i \rav=\lav 1-a_i-b_i \rav$.
When we consider a chain with open boundary conditions, particles are
injected and removed according to the boundary rates $\alpha_{A/B}, \gamma_{A/B},
\beta_{A/B}$ and $\delta_{A/B}$. So the attempt rate
for an $A$-particle to enter the system at the left boundary is $\alpha_A$. 
It leaves the channel at the left boundary with rate
$\gamma_A$ as illustrated in Fig.~\ref{latticeThreeState}. 
The other boundary rates are defined similarly.

In terms of a quantum Hamiltonian formalism \cite{Schu01} the probability 
distribution of the system is represented by the probability vector
$|P(t)>$ which evolves in time according 
to the master equation 
\begin{align}
 \frac{d}{dt}|P(t)>=-H|P(t)>
\end{align}
where the generator $H$ has the structure
\begin{align}
\label{hamil}
H=b_1+b_L+\sum_{i=1}^{L-1}h_{i,i+1}.
\end{align}
For an explicit representation of the generator we assign to
the three allowed single-site states the three basis vectors
\begin{align}
\label{basis}
|A>\equiv |1>=
\left(
  \begin{array}{l}
    1 \\
    0 \\
    0
  \end{array}
\right), \quad
|\emptyset>\equiv |0>=
\left(
  \begin{array}{l}
    0 \\
    1 \\
    0
  \end{array}
\right), \quad
|B>\equiv |-1>=
\left(
  \begin{array}{l}
    0 \\
    0 \\
    1
  \end{array}
\right)
\end{align}
corresponding to having an $A$, no particle or $B$ at site $i$, respectively.
A configuration of the full lattice is then represented by a tensor product 
of these basis vectors. To construct $H$, let $E_k^{x,y}$ be the 
$3 \times 3$ matrix with one element located at column $x$ and row $y$ 
equal to one. All other elements are zero: 
$(E_k^{x,y})_{a,b}=\delta_{x,a}\delta_{y,b}$. The operator for
annihilation (creation) of an $A$-particle at site $k$ is then
$a_k^-=E_k^{1,2}$ ($a_k^+=E_k^{2,1}$) and for annihilation (creation) of 
$B$ is $b_k^-=E_k^{3,2}$ ($b_k^+=E_k^{2,3}$). Finally, the number operators
are $a_k=E_k^{1,1}$, $b_k=E_k^{3,3}$, $v_k=1-a_k-b_k$. 
In this representation the boundary matrices are
\begin{align}
\label{boundary}
b_1=\alpha_A(v_1-a_1^+)+\alpha_B(v_1-b_1^+)+\gamma_A(a_1-a_1^-)+\gamma_B(b_1-b_1^-)\\
b_L=\delta_A(v_L-a_L^+)+\delta_B(v_L-b_L^+)+\beta_A(a_L-a_L^-)+\beta_B(b_L-b_L^-).
\end{align}
Hopping in the bulk between site $i$ and $i+1$ is generated by
\begin{multline}
\label{hopping}
h_{i,i+1}=D_A(a_iv_{i+1}+v_ia_{i+1} -a_i^-a_{i+1}^+ -a_i^+a_{i+1}^-) \\
      +D_B(b_iv_{i+1}+v_ib_{i+1} -b_i^+b_{i+1}^- -b_i^-b_{i+1}^+).
\end{multline}
The model is now well defined. We refer to appendix $A$ for 
more details on the formalism. 

\begin{figure}
\centerline{\includegraphics[width=14cm]{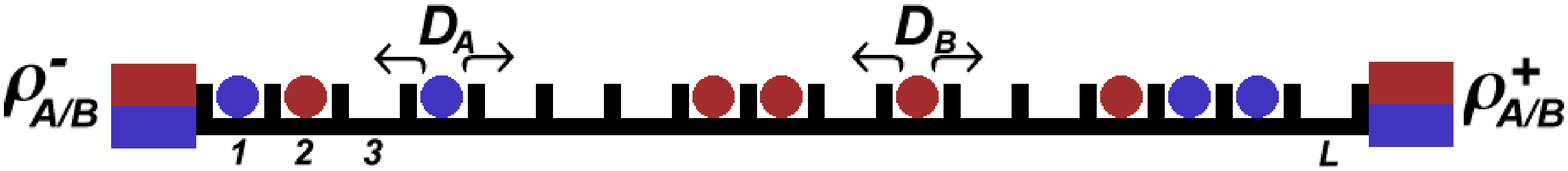}}
\caption{Three-state symmetric exclusion model with open boundaries -- reservoir picture.}
\label{latticeThreeStateReservoir}
\end{figure}

In \cite{Brza06} it was shown that with the choice
$\frac{\alpha_A}{\gamma_A}=\frac{\delta_A}{\beta_A}\equiv\mu_A$ and 
$\frac{\alpha_B}{\gamma_B}=\frac{\delta_B}{\beta_B}\equiv\mu_B$
the process has an uncorrelated stationary
equilibrium distribution with constant local particle densities
\begin{align}
  \label{densityA}
  \rho_A=
  \frac{\frac{\alpha_A}{\gamma_A}}{1+\frac{\alpha_A}{\gamma_A}+\frac{\alpha_B}{\gamma_B}}=
  \frac{\frac{\delta_A}{\beta_A}}{1+\frac{\delta_A}{\beta_A}+\frac{\delta_B}{\beta_B}}
\end{align}
\begin{align}
  \label{densityB}
  \rho_B=
  \frac{\frac{\alpha_B}{\gamma_B}}{1+\frac{\alpha_A}{\gamma_A}+\frac{\alpha_B}{\gamma_B}}=
  \frac{\frac{\delta_B}{\beta_B}}{1+\frac{\delta_A}{\beta_A}+\frac{\delta_B}{\beta_B}}.
\end{align}
Besides giving the equilibrium densities, these two equations above provide 
a recipe of how to translate the picture of inserting and deleting particles 
on boundary sites into a picture of constant reservoirs at the ends, see
Fig.~\ref{latticeThreeStateReservoir}. One may regard the middle expression
in (\ref{densityA}) as a left reservoir density $\rho_A^-$ of $A$-particles,
located at a virtual site $i=0$, with which the system exchanges 
particles by jump events with rates $D_{A,B}$. Similarly, the middle expression
in (\ref{densityB}) is a regarded as left reservoir density $\rho_B^-$ of 
$B$-particles. At the right boundary one adds a virtual reservoir
(located at site $i=L+1$) with boundary densities $\rho_A^+$, $\rho_B^+$
given by the rightmost expressions in (\ref{densityA}), (\ref{densityB}).

In this picture, jumping from a reservoir site onto the boundary of the system occurs
proportional to the respective hopping rate and proportional to the single 
species boundary density. This parameterisation satisfies \eqref{densityA} 
and \eqref{densityB} if
\begin{align}
\label{boundaryratesleft}
\alpha_{A/B}=D_{A/B}\rho_{A/B}^-, \quad
\gamma_{A/B}=D_{A/B}(1-\rho_A^--\rho_B^-)
\end{align}
\begin{align}
\label{boundaryrateslright}
\delta_{A/B}=D_{A/B}\rho_{A/B}^+, \quad
\beta_{A/B}=D_{A/B}(1-\rho_A^+-\rho_B^+).
\end{align}
The second equality 
in (\ref{densityA}), (\ref{densityB}) is then the equilibrium condition
$\rho_{A,B}^-=\rho_{A,B}^+$ of equal reservoir densities or chemical
potentials $\mu_{A,B}^-=\mu_{A,B}^+$. In a non-equilibrium setting the 
left reservoir densities  $\rho_A^-$, $\rho_B^-$ are not equal to those on 
the right boundary ($\rho_A^+$, $\rho_B^+$). In this boundary-driven case 
the system evolves towards a stationary state with non-vanishing particle 
currents.

Notice that for periodic boundary conditions the system is highly
non-ergodic not only because of particle number conservation, but also
because of the single-file constraint which conserves the sequence
of particles. The open system, on the other hand, is ergodic 
because of the violation of the conservation of 
particles at the boundaries.

\section{Relaxation}

\subsection{Hydrodynamic limit}

The average densities $\lav a_i \rav$ and $\lav b_i \rav$ satisfy the 
equations of motion
$\frac{d}{dt}\lav a_i \rav=-\lav a_i H \rav$, 
$\frac{d}{dt}\lav b_i \rav=-\lav b_i H \rav$. 
This provides the Master equations for a single site,
\begin{align}
\label{motion}
\frac{d}{dt}\lav a_i \rav
&= D_A\left( \lav a_{i-1}v_i \rav + \lav a_{i+1}v_i \rav - 
\lav a_i v_{i+1} \rav - \lav a_i v_{i-1} \rav \right)\\
\frac{d}{dt}\lav b_i \rav
&= D_B\left( \lav b_{i-1}v_i \rav + \lav b_{i+1}v_i \rav - 
\lav b_i v_{i+1} \rav - \lav b_i v_{i-1} \rav \right).
\end{align}
At this stage we assume an infinite system and do not care about boundary 
sites. Nevertheless, in this form the equations of motion are not integrable. 
Replacing the joint probabilities by products of expectation values -- 
according to a simple mean field ansatz which has been proven useful in other
exclusion processes -- fails to provide an even qualitatively correct 
description. However, an exact equation containing no correlators can be 
obtained from a weighted sum
\begin{align}
\label{heat}
\frac{d}{dt}\left( \frac{\lav a_i \rav}{D_A} + \frac{\lav b_i \rav}{D_B} \right)
=\left( \lav a_{i-1} \rav + \lav a_{i+1} \rav - 2\lav a_i \rav + 
\lav b_{i-1} \rav + \lav b_{i+1} \rav - 2\lav b_i \rav \right).
\end{align}
of the individual particle densities. 

The right-hand side contains a second-order difference for both species 
individually. Coarse-graining, i.e., replacing $i$ by the continuous 
variable $x=\frac{i}{a}$ and introducing $\rho_A(x,t)$, $\rho_B(x,t)$
as the $A$, $B$-particle density at $x$, transforms \eqref{heat} 
(for vanishing lattice constant $a$) into the macroscopic equation
\begin{align}
\label{heatlimit}
\partial_t \left( \frac{\rho_A(x,t)}{D_A} + \frac{\rho_B(x,t)}{D_B} \right)
=\partial_x^2(\rho_A(x,t)+\rho_B(x,t)).
\end{align}
Here we used diffusive rescaling of the time-coordinate. Notice for the
hydrodynamic limit \eqref{heatlimit} to be valid we
implicitly use good local ergodic properties in the identification
of the lattice expectation with the coarse-grained deterministc particle 
density. The weighted density on the l.h.s. of \eqref{heatlimit}
and the total density on the r.h.s.
reappear in the computations below and for notational simplicity
we introduce 
\begin{align}
\label{sigma}
\rho(x,t) = \rho_A(x,t) + \rho_B(x,t)
\sigma(x,t) = \frac{\rho_A(x,t)}{D_A} + \frac{\rho_B(x,t)}{D_B}
\end{align}
which reduces to $\sigma = \rho/D$ for equal single-particle diffusion
coefficients $D_A=D_B\equiv D$.

The linear equation \eqref{sigma} does not provide a full description of the
coarse-grained dynamics. To achieve this goal we generalize the approach of 
\cite{Quas92} and make an ansatz for the dynamics of a single particle 
(of species $A$ or $B$) localized at position $x$. This test particle 
acts as a tracer particle in the background of other particles,
with a diffusive motion partially determined by its self-diffusion 
coefficient. Additionally, the test particle is subject to a background
drift $b$ caused by the collective evolution of the entire system towards 
stationarity. A priori the self-diffusion coefficient for the two species of 
particles could be different. However, due to the single-file condition the
long-time dynamics of a tracer particle is entirely determined by the
background of other particles which is the same for both species. Hence
we expect identical self-diffusion coefficients $D_s$ \cite{Asla00}. This physical
picture leads us to the ansatz
\begin{align}
\partial_t\rho_A(x,t)&=\partial_x^2 D_{s}\rho_A(x,t)-
\partial_x b(x,t)\rho_A(x,t) \label{ansatz1}\\
\partial_t\rho_B(x,t)&=\partial_x^2 D_{s}\rho_B(x,t)-
\partial_x b(x,t)\rho_B(x,t). \label{ansatz2}
\end{align}
for the time evolution of the coarse grained particle densities.
The drift term $b$ can then be determined by using \eqref{heatlimit}
\begin{align}
\label{theb}
b=\frac{1}{\sigma} \partial_x (D_{s} \sigma -\rho).
\end{align}

In an infinite system the self-diffusion coefficient vanishes, as
is indicated by the subdiffusive nature of single-file diffusion which was proved
rigorously for tracer diffusion in the SEP in \cite{Arra83}. However, as
argued in \cite{Brza06} we expect that in a finite system with open boundaries
correction terms of leading order $1/L$ appear. This is confirmed by the 
exact result 
\begin{align}
\label{ds}
D_s=\frac{1}{L}\frac{1-\rho}{\sigma}
\end{align}
derived in Appendix \ref{selfdiff} for a periodic system.
Introducing the two-component local order parameter
\begin{align}
\label{vector}
\vec{\rho}=
\left(
  \begin{array}{l}
    \rho_A \\
    \rho_B
  \end{array}
\right)
\end{align}
finally leads to the coupled nonlinear diffusion equation
\begin{align}
\label{hydro}
\partial_t \vec{\rho} = \partial_x \left( \hat{D}  \partial_x \vec{\rho} \right)
\end{align}
with the diffusion matrix
\begin{align}
\label{matrix}
\hat{D} = \frac{1}{\sigma} \left[
\left(
  \begin{array}{cc}
    \rho_A & \rho_A\\
    \rho_B & \rho_B
  \end{array}
\right) 
+ D_s \left(
  \begin{array}{cc}
    \frac{\rho_B}{D_B} & -\frac{\rho_A}{D_B} \\
    -\frac{\rho_B}{D_A} & \frac{\rho_A}{D_A}
  \end{array}
\right) \right].
\end{align}
This equation
provides a complete description of the density dynamics in the hydrodynamic
limit. For given initial density profiles the solution of \eqref{ansatz1}, 
\eqref{ansatz2} is uniquely deternined. Notice that for an infinite
system where $D_s=0$ the diffusion matrix $\hat{D}$ has a vanishing 
eigenvalue. For $D_A=D_B$ \eqref{ansatz1} and \eqref{ansatz2} reduce to 
the coupled nonlinear diffusion equations proved rigorously in 
\cite{Quas92}.

\subsection{Time evolution of density profiles}

\begin{figure}
 \includegraphics[width=5.5cm,angle=270]{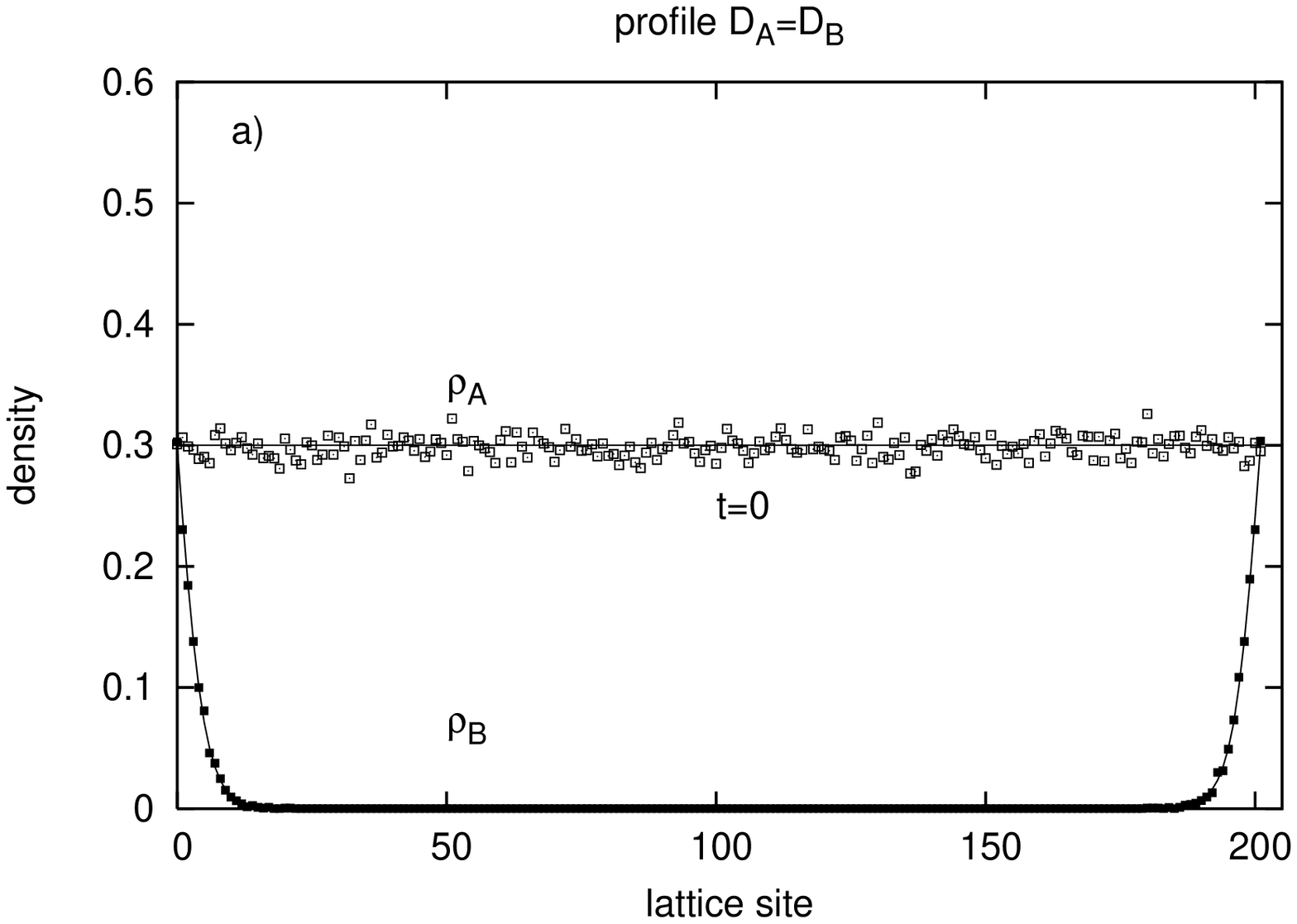}
 \includegraphics[width=5.5cm,angle=270]{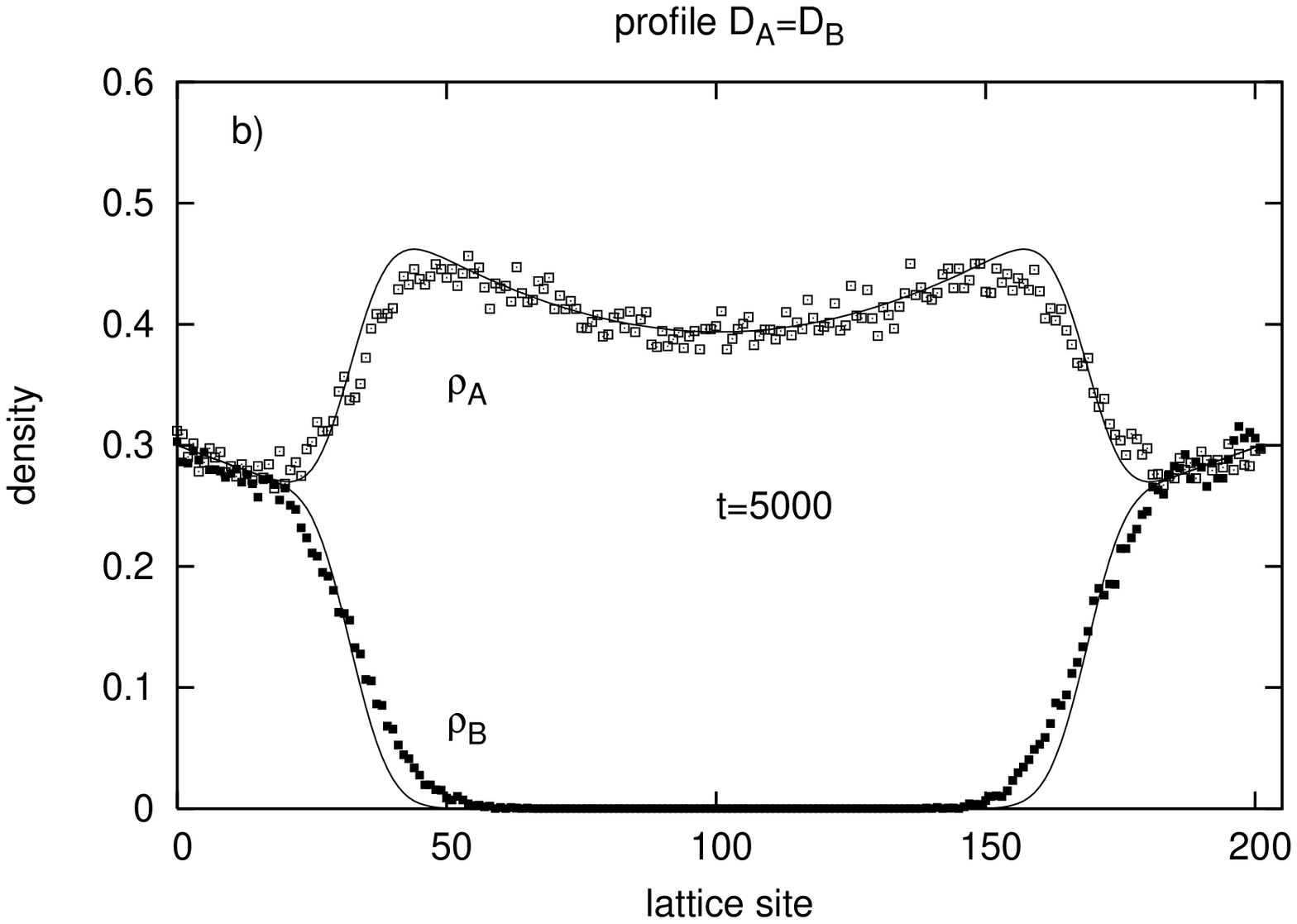}
 \includegraphics[width=5.5cm,angle=270]{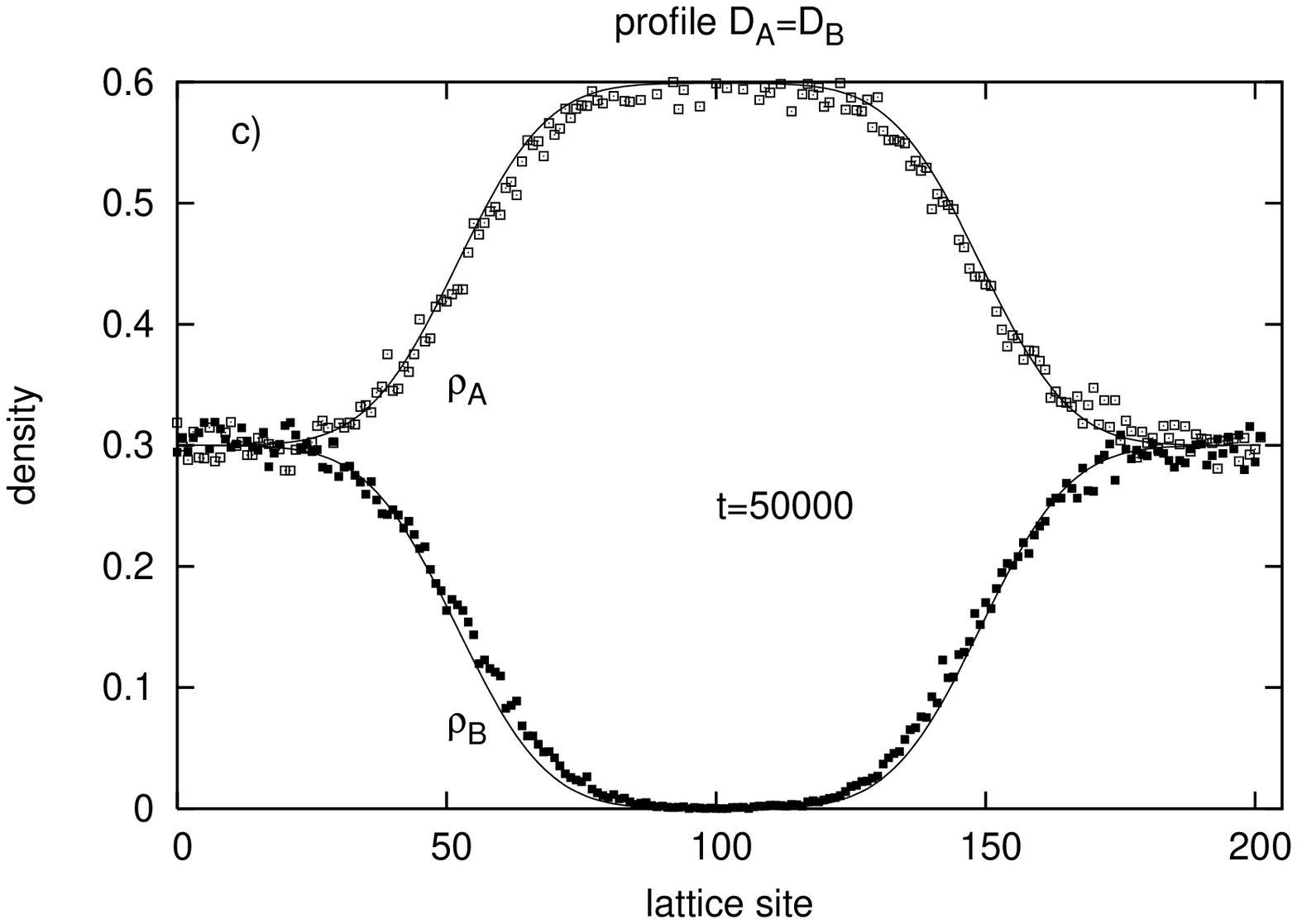}
 \includegraphics[width=5.5cm,angle=270]{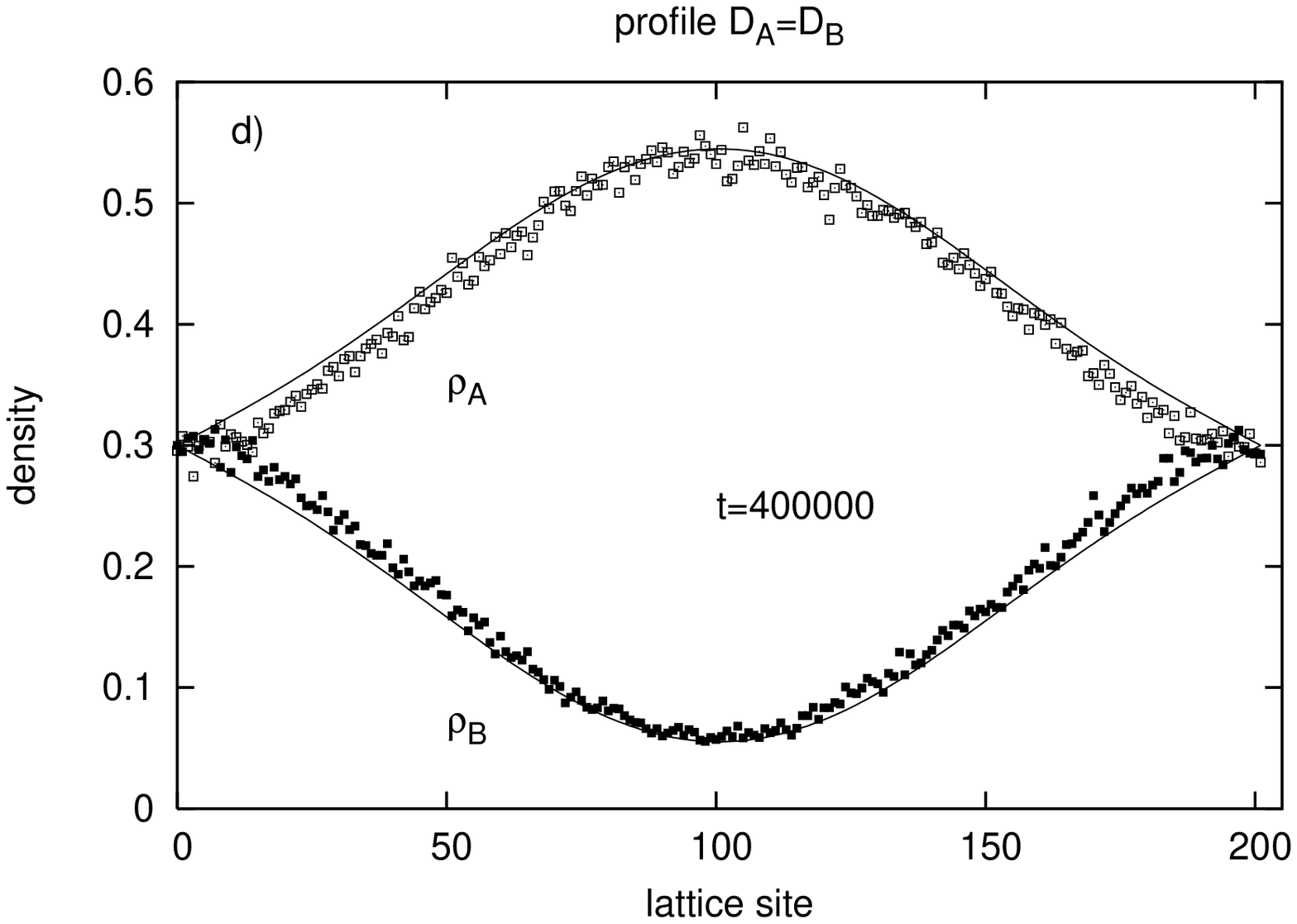}
\caption{Density profiles of the $A$($B$)-component
(upper/lower curve) at different MC times for a system ($L=200$)
with equal diffusion coefficients. Snapshots are taken at a) $t=0$, b) $t=5000$, c) $t=50000$, d) $t=400000$.
The system relaxes towards the equilibrium state given by the reservoir
densities $\rho_{A/B}^{+/-}=0.3$. Symbols denote densities obtained from DMCS. 
Full lines come from numerical integration of the diffusion equation.}
\label{tracer}
\end{figure}

We have no analytical expression for the solution of the system \eqref{hydro}, 
but numerical integration can be performed to great accuracy using standard 
routines. We first study the case of equal hopping rates $D_A=D_B$. This case is 
similar to the tracer diffusion investigated in \cite{Kaerg2002}. Fig. 
\ref{tracer} shows the penetration of tracer particles into the
system with lattice size $L=200$ at 
four different times. Full lines are solutions of \eqref{hydro} obtained 
from numerical integration. Circles indicate the numerical DMCS densities.
Initially, the lattice is prepared with a uniform distribution of $B$-particles
and very few $A$-particles. The smooth drop of the initial $A$-particle profile 
at the edges was chosen for faster numerical integration. The boundary 
reservoir densities are $\rho_A^-=\rho_A^+=\rho_B^-=\rho_B^+=0.3$. For all 
instances of time the numerical integration shows a rather satisfying 
collapse with the profiles obtained from Monte Carlo simulation.
Any DMCS in this work was averaged over at least 
10000 realizations.

It is remarkable that the PDEs describe the intermediate flattening at the 
boundary (see e.g. Fig.~\ref{tracer}c) of the $A$-particle profile which was 
observed for the first time in \cite{Kaerg2002}. It is also interesting that 
the spatial mean density of $B$-particles first grows before it relaxes to its 
equilibrium value 0.3. The explanation comes from the hydrodynamic equation 
\eqref{heatlimit} which states that the total density profile, i.e. the sum of 
both species, evolves according to an ordinary diffusion equation. Hence, the
relaxation to equilibrium is proportional to $L^2$, in contrast to the much
slower relaxation time of order $L^3$ of the individual components. The 
$B$-particles compensate the $A$-particle profile until both reach equilibrium.

\begin{figure}
 \includegraphics[width=5.5cm,angle=270]{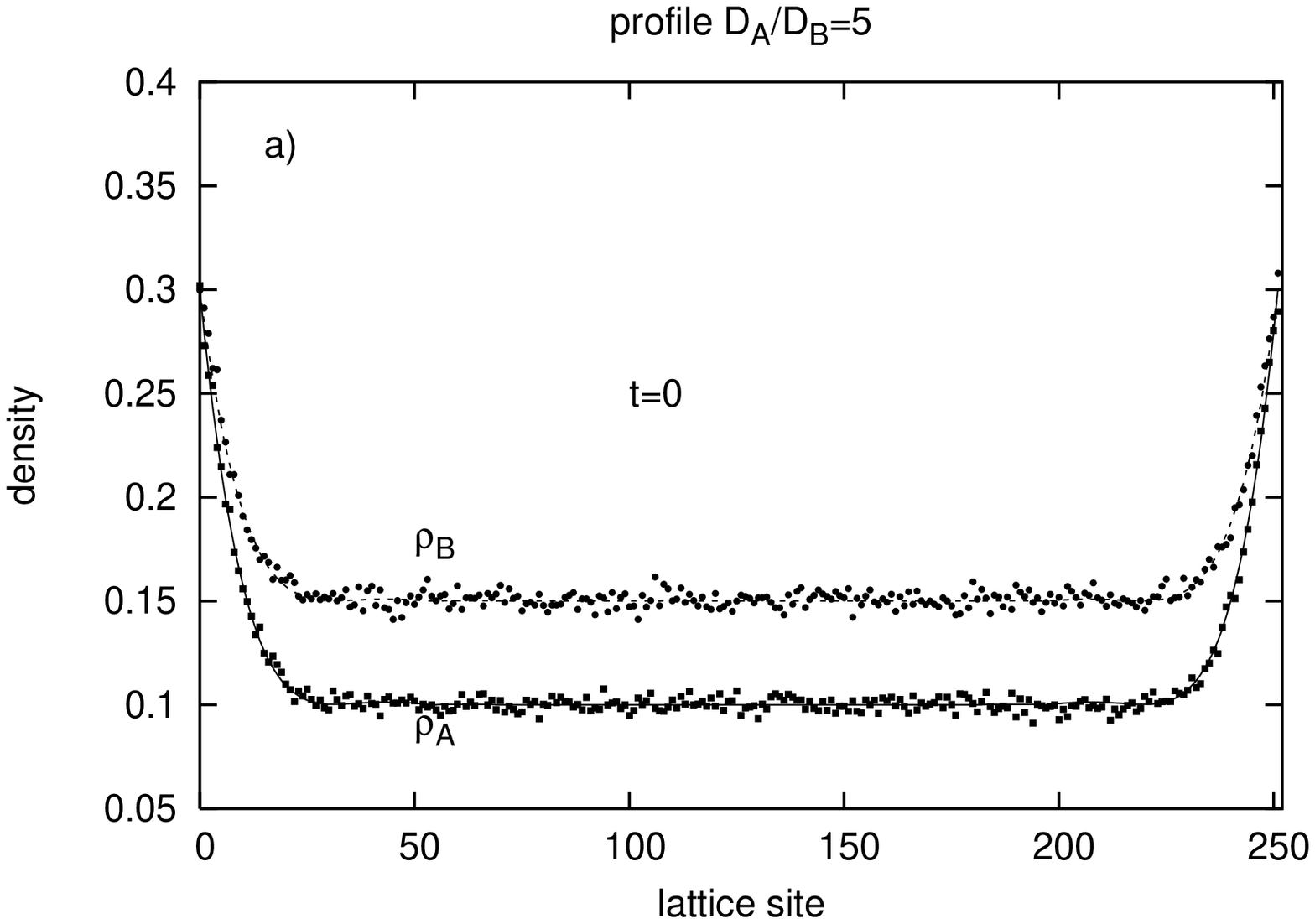}
 \includegraphics[width=5.5cm,angle=270]{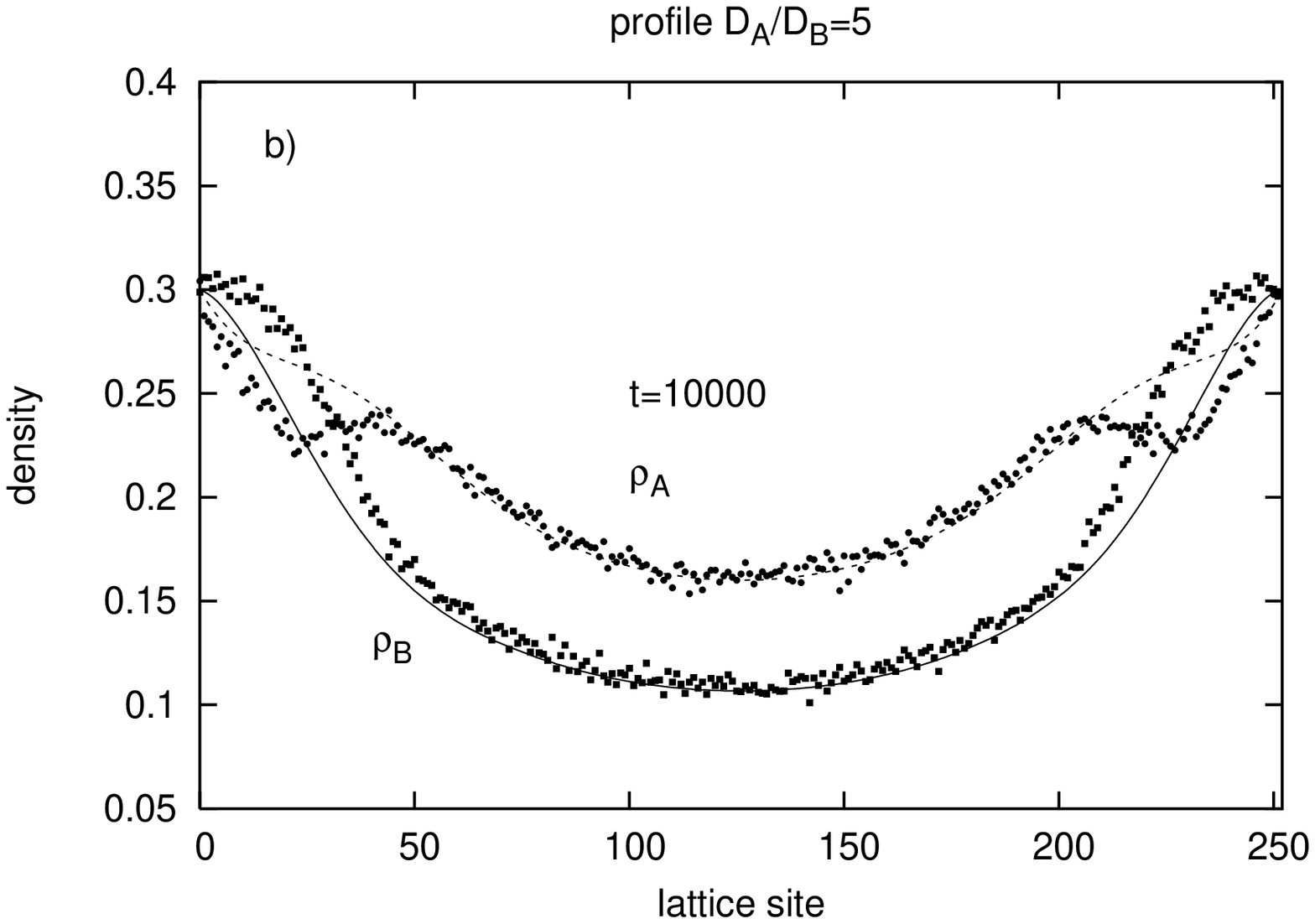}
 \includegraphics[width=5.5cm,angle=270]{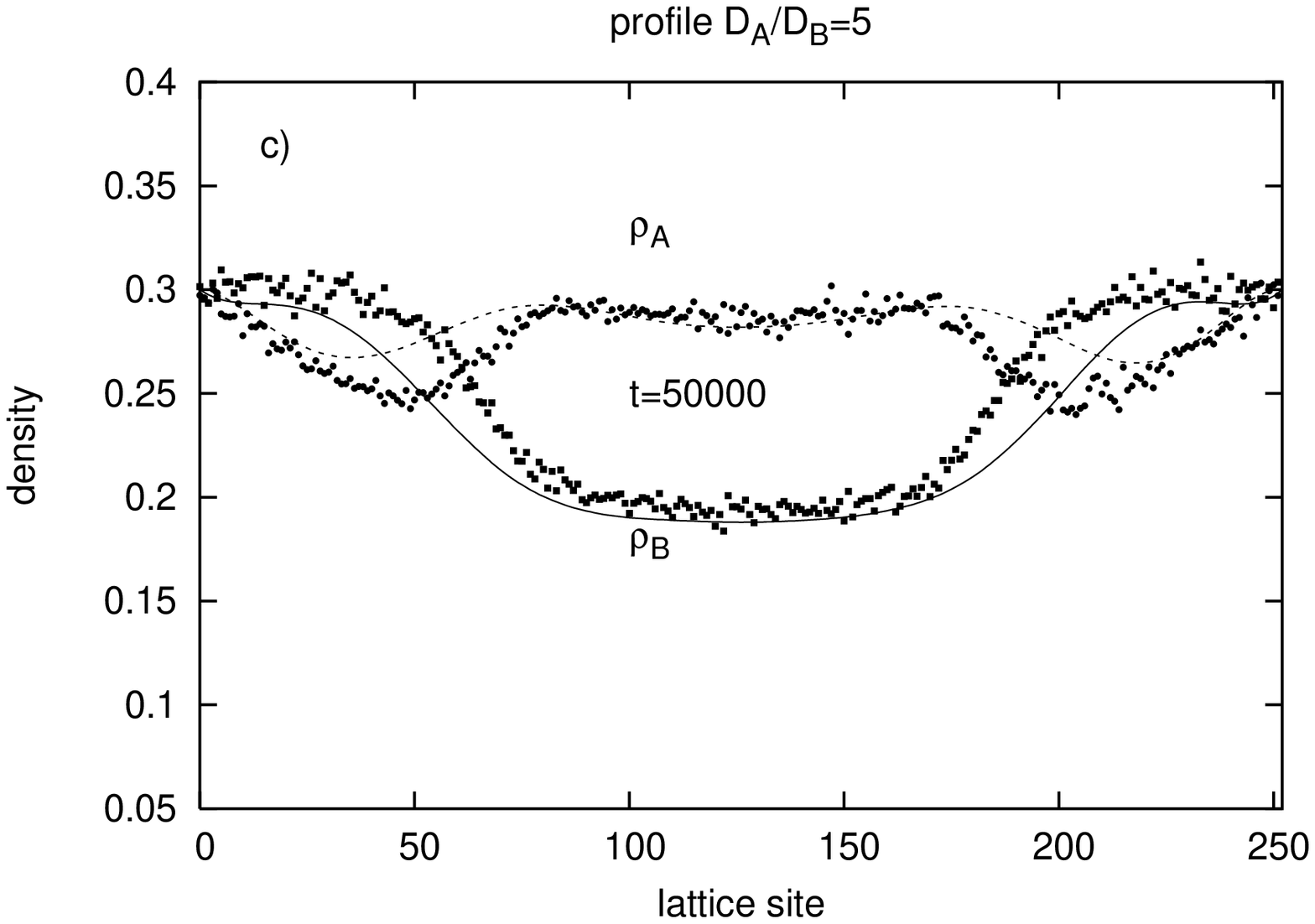}
 \includegraphics[width=5.5cm,angle=270]{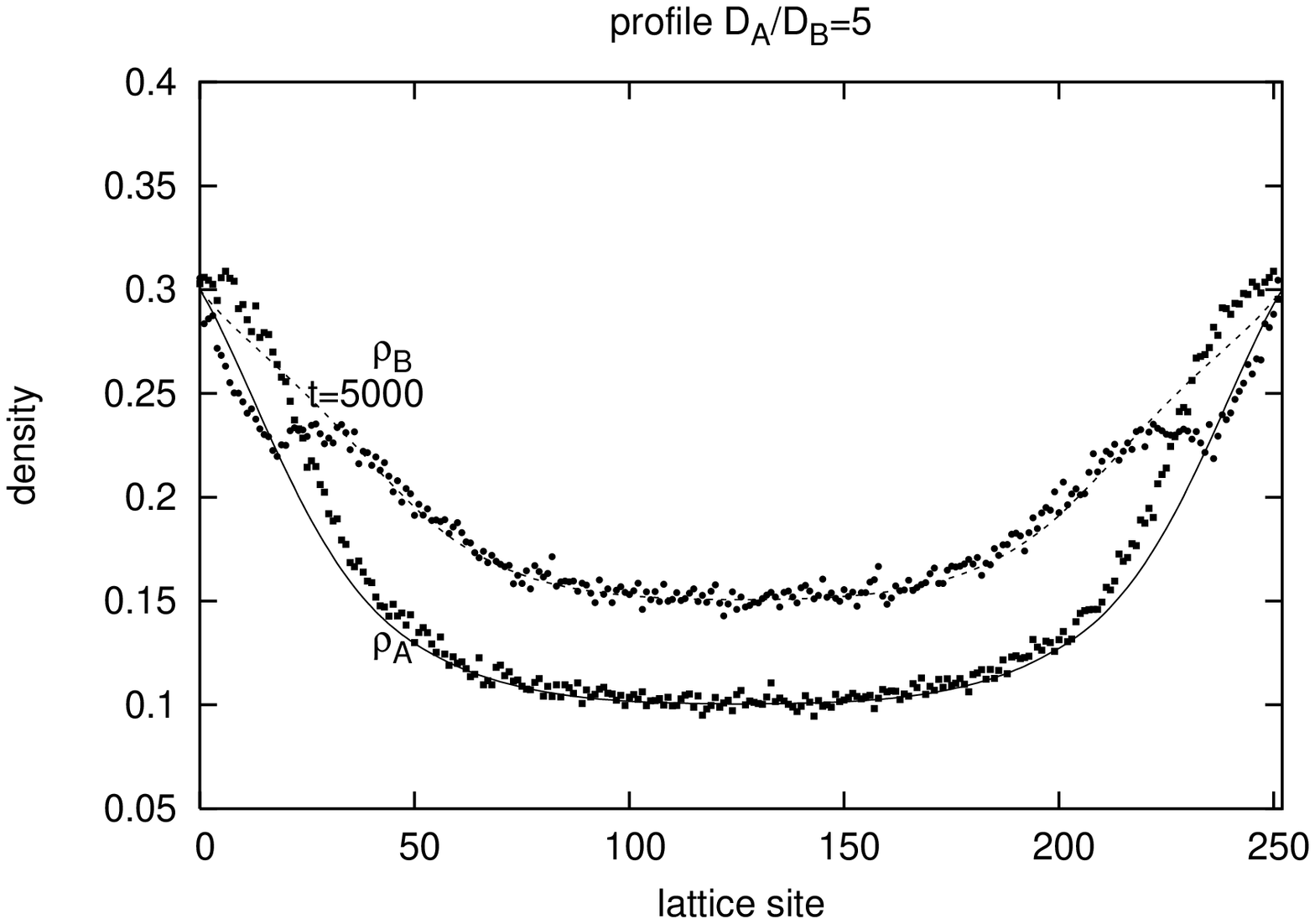}
\caption{Snapshots of the density profiles for a system with $L=250$, 
$D_A/D_B=5$ and open boundary conditions with reservoir
densities $\rho_{A/B}^{+/-}=0.3$. MC times are a) $t=0$, b) 
$t=10000$, c) $t=50000$. Circles denote densities obtained from DMCS. 
Full lines come from numerical integration of the hydrodynamic equation. 
The stronger discrepancy between simulation data and integration shown in 
Fig. d) for $L=100$ and $t=5000$ demonstrates significant 
finite size effects.}
\label{openbound}
\end{figure}

As a second example we consider the relaxation of a mixed system with
heavy $B$ particles and light $A$-particles where $D_A/D_B=5$.
For Fig.~\ref{openbound} a)-c) we have chosen a model with $L=250$, 
and reservoir densities $\rho_A^-=\rho_A^+=\rho_B^-=\rho_B^+=0.3$ as
before. The agreement between simulation and integration in the vicinity 
of the boundary is not as good as for tracer particles. Nevertheless, 
comparing the simulation results for the smaller lattice ($L=100$) 
shown in d) with the results for the larger lattice shown in c) 
suggests that the deviations are finite-size effects. We do not have 
a theory of finite-size effects beyond the leading $1/L$-correction 
to the self-diffusion coefficient. Therefore we do not pursue this 
issue further.

\subsection{Small diffusion rate for one of the species}

We discuss the case where one of the species, say $B$, is very heavy 
and has a vanishing hopping rate $D_B \to 0$. In this case
even the open system is not ergodic and
the stationary state to which the system relaxes depends on
the initial distribution of $A$ and $B$-particles. $B$-particles just remain
at their initial position and the equilibrium $A$-particle density 
between two $B$'s is just their local initial density. 
Let us now assume an initial state without any $B$-particles in the channel 
and let the reservoir densities be non-vanishing with 
$\rho_A=\rho_A^-=\rho_A^+$ and
$\rho_B=\rho_B^-=\rho_B^+$. For the case of strictly vanishing
hopping rate $D_B=0$ both injection rates of $B$-particles at 
the boundaries vanish
and therefore no $B$-particles ever enter the system.
Then the bulk dynamics follow the rules of the single-species SEP, but
an anomaly comes from the fact that injection of $A$-particles is proportional
to the $A$-particle density whereas the outgoing rate occurs according
to the hole density which involves the reservoir density $\rho_B\neq 0$. Hence,
$A$-particles do not equilibrate towards $\rho_A$ 
(as expected from the single-species SEP) but reach a higher value:
\begin{align}
\label{exceed}
\rho^{D_B=0}_A=\frac{\alpha_A}{\alpha_A+\gamma_A}=
\frac{\delta_A}{\delta_A+\beta_A}=\frac{\rho_A}{1-\rho_B}.
\end{align}
Notice that any ratio $\rho^{D_B=0}_A/\rho_A>1$ can be 
achieved provided $\rho_B$ and $D_A$ are
large enough. 

\begin{figure}
\centerline{\includegraphics[width=9cm,angle=270]{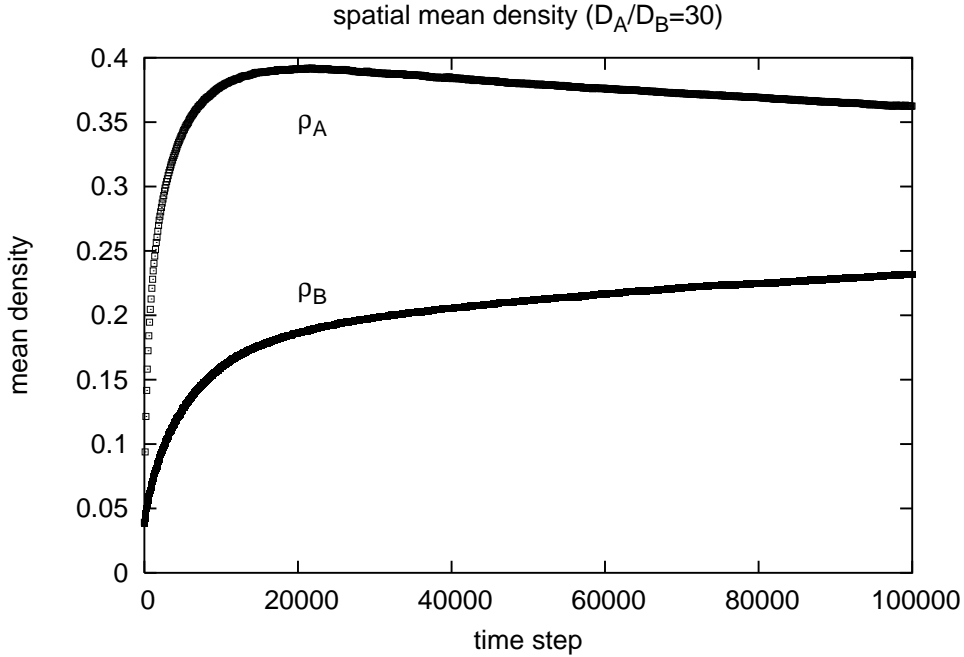}}
\caption{Mean density of $A$ and $B$-particles as a function of MC steps. The relaxation
towards the equilibrium values $\rho_{A/B}^{+/-}=0.3$ was started from an almost
empty lattice ($L=50$).}
\label{overshooting}
\end{figure}

This leads to an interesting conclusion for small but non-vanishing $0<D_B\ll D_A/L^2$.
Then channel is in a quasi-equilibrium with $A$-particles (with density \eqref{exceed})
before eventually a $B$-particle enters the system.  Only very slowly both species
start to relax to their common equilibrium state and one expects the $A$ 
profile to initially exceed its true equilibrium value $\rho_A$, up to some
maximal value which is bounded by the maximal density 
\eqref{exceed}. This overshooting is indeed observed in simulation, 
see Fig.~\ref{overshooting} for the time evolution of the spatial mean densities. 
Initially the system ($L=50$) is almost empty and the particles
are injected and removed at the boundaries according to the reservoir densities
$\rho^{+/-}_{A/B}=0.3$. The ratio of the diffusion coefficient is $D_A/D_B=30$. One 
can clearly see the excess of $A$-particles which disappears only very slowly. 

\section{Steady state behaviour}

\subsection{Finite $L$}

In the steady state the time derivative in the diffusion equation \eqref{hydro}
vanishes and from \eqref{heatlimit} we obtain the integration constant
\begin{align}
\partial_x \rho = - c.
\end{align}
In terms of the individual stationary particle currents
we have $c = j_A/D_A + j_B/D_B$.
Integration yields the stationary total density profile $\rho=\rho_A+\rho_B$
\begin{align}
\label{totaldensity}
\rho(x) = \rho^- + \frac{\rho^+ - \rho^-}{L} x
\end{align}
in terms of the boundary values $\rho^\pm$. In terms of the boundary 
densities the integration constant is therefore given by
\begin{align}
\label{totalcurrent}
c = - \frac{\rho^+ - \rho^-}{L}.
\end{align}

In order to compute the density profile of the $A$-species we introduce
\begin{align}
h(x) \equiv \frac{\rho_A(x)}{\sigma(x)} = \frac{D_B\rho_A(x)}{\rho(x) - \left(1-\frac{D_B}{D_A}\right)\rho_A(x)}.
\end{align}
Given $h$ both $\rho_A$ and $\rho_B$ can be computed from \eqref{totaldensity}.
Using the self-diffusion coefficient \eqref{ds}, the stationarity condition for $\rho_A$ becomes
\begin{align}
\frac{1}{L} \partial_x \left[ (1-\rho)h \right] + \left(1 + \frac{1}{L} \right) h \partial_x \rho = - j_A
\end{align}
with an integration constant $j_A$ proportional to the current of $A$-particles. Using the
linearity of the total density profile this yields
\begin{align}
\label{stationarity}
\frac{1-\rho(x)}{L} h'(x) -c h(x) + j_A = 0.
\end{align}
where the prime denotes the derivative w.r.t. $x$.

This differential equation is straightforwardly integrated and one
finds
 \begin{align}
\label{stationaryh}
h(x) = h^- - \frac{h^- - h^+}{1 - \left(\frac{1-\rho^+}{1-\rho^-}\right)^L}
\left[1 - \left(1- \frac{\rho^+-\rho^-}{1 -\rho^-} \frac{x}{L}\right)^L\right].
\end{align}
Here we have introduced the
boundary value
\begin{align}
\label{boundaryh}
h^\pm =  \frac{\rho_A^\pm}{\frac{\rho_A^\pm}{D_A} + \frac{\rho_B^\pm}{D_B}}.
\end{align}
Solving for the density profile of $A$-particle then yields
 \begin{align}
\label{stationaryrhoA}
\rho_A(x) = \left[\rho^- + (\rho^+-\rho^-) \frac{x}{L}\right]
\frac{h(x)}{D_B + \left(1-\frac{D_B}{D_A}\right)h(x)}.
\end{align}
The integration constant $j_A$ is given in terms of boundary densities by
\begin{align}
\label{stationaryja}
j_A = \frac{ h^+ \left(1-\rho^-\right)^L 
           - h^- \left(1-\rho^+\right)^L  }
         {  \left(1-\rho^-\right)^L 
          - \left(1-\rho^+\right)^L  } \frac{\rho^- - \rho^+}{L}.
\end{align}
Generically this quantity is of order $1/L$, up to corrections which are
exponentially small in system size as long as $\rho^-\neq\rho^+$.

For vanishing reservoir gradient $\rho^+ = \rho^- \equiv \rho$ we have $c=0$.
This yields a linear function
\begin{align}
\label{stationaryh2}
h(x) = h^- - (h^- - h^+) \frac{x}{L}
\end{align}
and nonlinear density profile
\begin{align}
\label{stationaryrhoA2}
\rho_A(x) = \frac{ \rho_A^-\left[1- \left(1-\frac{D_B}{D_A}\right) \frac{\rho_A^+}{\rho}\right] 
                    + \left(\rho_A^+ - \rho_A^-\right) \frac{x}{L}}
                 {1- \left(1-\frac{D_B}{D_A}\right) \frac{\rho_A^+}{\rho}
                    +\left(1-\frac{D_B}{D_A}\right) \frac{\rho_A^+-\rho_A^-}{\rho} \frac{x}{L}}.
\end{align}
The current of $A$-particles
\begin{align}
\label{stationaryja2}
j_A = \frac{1-\rho}{L^2} 
\left( h^- - h^+ \right)
\end{align}
is only of order $1/L^2$, as opposed to the generic $1/L$ dependence.
For the total particle current we find
\begin{align}
\label{stationarycurrent}
j \equiv j_A+j_B =  \frac{\rho(1-\rho)}{L^2} 
 \frac{\left(\rho_A^+ - \rho_A^- \right)\left(D_A^{-1}-D_B^{-1}\right)}{\sigma^+\sigma^-} = \frac{\rho}{L} (D_s^- - D_s^+)
\end{align}
which is also of order $1/L^2$.
We remark that this current is proportional to the boundary gradient of 
the
self-diffusivity.

\subsection{Phase transition in the thermodynamic limit}

The expression for the $A$-particle density derived for finite $L$ is
cumbersome. Rather interesting behaviour emerges from an analysis of the
thermodynamic limit $L\to\infty$. To leading order in $1/L$ we find
\begin{align}
\label{limitja}
j_A = \left\{ \begin{array}{ll}
    \frac{\rho_A^+}{\frac{\rho_A^+}{D_A} + \frac{\rho_B^+}{D_B}}  \frac{\rho^- - \rho^+}{L} & \mbox{ for } \rho^- < \rho^+ \\
    \frac{\rho_A^-}{\frac{\rho_A^-}{D_A} + \frac{\rho_B^-}{D_B}}  \frac{\rho^- - \rho^+}{L} & \mbox{ for } \rho^- > \rho^+
  \end{array}
 \right.
\end{align}
Hence, as one expects, the $A$-current is always opposite the total reservoir gradient $\Delta \rho \equiv  \rho^+ - \rho^-$, but it changes in a 
non-analytic fashion at $\rho^- = \rho^+$ where 
it vanishes to leading order in $1/L$, see \eqref{stationaryja2}. 
For positive reservoir gradient $\Delta \rho$ the amplitude of $A$-current is 
governed by the reservoir density at right boundary, otherwise by the left boundary.

A nonanalyticity at $\Delta \rho=0$ appears also in the behaviour of the 
density. Using $\lim_{L\to\infty} (1-ax/L)^L = {\rm e}^{-ax}$ yields the following behaviour.

\noindent \underline{\it Case 1, $\rho^- < \rho^+$:} 
Straightforward algebra turns \eqref{stationaryh} into
\begin{align}
\label{asymh}
h(x) = h^+ + (h^--h^+){\rm e}^{-x/\xi}
\end{align}
with localization length
\begin{align}
\label{localization}
\xi = \left[ \ln{\frac{\rho^+ - \rho^-}{1-\rho^-}} \right]^{-1}.
\end{align}
This gives
 \begin{align}
\label{stationaryrhoAlimit}
\rho_A(x) = \left[\rho^- + (\rho^+-\rho^-) \frac{x}{L}\right]
\frac{h^+ + (h^--h^+) {\rm e}^{-x/\xi} }
     {D_B + \left(1-\frac{D_B}{D_A}\right)\left[h^+ + (h^--h^+) {\rm e}^{-x/\xi}\right]}.
\end{align}

For large distance $x\gg \xi$ from the left boundary the function $h$ 
approaches a constant and the density profile of $A$-particles becomes 
proportional to the total density
\begin{align}
\label{aprofile}
\rho_A(x) \to \frac{\rho_A^+}{\rho^+} \rho(x).
\end{align}
Therefore at large distance from the boundary the density of each
species becomes a linear function. The exponentially decaying
deviation from linearity marks the occurrence of a boundary layer
of width $\xi$ at the left boundary. Boundary layers in two-component
systems with open boundaries were previously observed in
\cite{Brza06,Vase06} and in earlier work \cite{Kaerg2002,Leeuwen2003}.
The asymptotic space-averaged mean density 
\begin{align}
\bar{\rho}_A = \lim_{L\to\infty}\frac{1}{L} \int_0^L dx \rho_A(x)
\end{align}
takes the value
\begin{align}
\label{aaverage}
\bar{\rho}_A =  \frac{\rho_A^+}{\rho^+} \frac{\rho^++\rho^-}{2}
\end{align}
which is independent of $D_B/D_A$.

\noindent \underline{\it Case 2, $\rho^- > \rho^+$:} 
The computation for negative reservoir gradient $\Delta\rho <0$ is analogous
and all results can be obtained from Case 1 by interchanging $(+,-)$ and
$(x,L-x)$.
One finds
\begin{align}
\label{asymh2}
h(x) = h^- - (h^--h^+){\rm e}^{-(L-x)/\xi}
\end{align}
with localization length
\begin{align}
\label{localization2}
\xi =\left[ \ln{\frac{\rho^- - \rho^+}{1-\rho^+}} \right]^{-1}. 
\end{align}
This corresponds to a boundary layer at the right boundary of the system.
For large distance $L-x\gg \xi$ from the right boundary the function $h$ 
approaches a constant and the
the density profile of $A$-particles becomes proportional to the
total density
\begin{align}
\label{aprofile2}
\rho_A(x) \to \frac{\rho_A^-}{\rho^-} \rho(x)
\end{align}
Correspondingly one has
\begin{align}
\label{aaverage2}
\bar{\rho}_A =  \frac{\rho_A^-}{\rho^-} \frac{\rho^++\rho^-}{2}
\end{align}
Therefore
the mean density as a function of the reservoir gradient $\Delta\rho$
has a jump discontinuity at $\Delta\rho=0$. In each of the two
phases it is controlled by the boundary which does not exhibit the
boundary layer.

\section{Conclusions}

The paper is dedicated to a detailed quantitative investigation 
single-file diffusion with two species of particles in an open system. 
Adapting ideas from the probabilistic hydrodynamic approach and 
using information from the microscopic master equation lead us to 
derive a nonlinear diffusion equations for the evolution of the 
macroscopic particle densities. We compute the exact self-diffusion 
coefficients $D_s$ for the two species which are equal and of order $1/L$.
Hence they vanish in the hydrodynamic limit, in agreement with the
subdiffusive behaviour of single-file diffusion. However, setting $D_s=0$
renders the diffusion matrix singular (with a vanishing eigenvalue)
and the boundary-value problem becomes overdetermined. Therefore
we conclude that $D_s$ is required for regularization. By use of
numerical integration and Monte Carlo simulation the
relaxation for the two species towards equilibrium is analysed.
The very slow relaxation of the individual particle species is affected 
by a fast relaxation of the sum 
of both species. The density of the fast species reaches a maximum 
before it starts to relax slowly to its actual equilibrium value.
From the diffusive nature of the collective behaviour we conclude
that the collective relaxation time is of order $L^2$. The 
$1/L$-behaviour of the self-diffusion coefficient then implies
single-species relaxation times of order $L^3$.

The diffusion equations are solved analytically for the stationary case. 
There is a boundary-induced non-equilibrium phase transition of first order.
This transition resembles a similar transition observed in previous work 
\cite{Brza06}
in so far as (i) this transition occurs when there is no gradient in the 
reservoir densities $\rho^+,\rho^-$ and (ii) the transition
is accompanied by a 
discontinuous change of the position of the boundary layer from
one boundary to the other. 
However, a detailed analysis of our results
reveals some interesting new features. Neither the width of the 
boundary layer 
nor the stationary bulk density profiles \eqref{aprofile} depend on the ratio $D_B/D_A$
inside the two phases.
At the
phase transition line $\Delta\rho=0$, however, the individual density profiles 
are strongly sensitive to the ratio $D_B/D_A$. Unlike in the case of equal hopping
rates the total particle current $j$ does {\it not} vanish, even though 
the total density $\rho = \rho_A+\rho_B$ is constant and equal to the reservoir densities.
In terms of currents the phase transition occurs when the weighted current
$c = j_A/D_A + j_B/D_B$ vanishes.

Boundary layers are known to occur also in bulk-driven lattice gases and have
been analyzed in some detail in a series of recent papers 
\cite{Anta00,Popk03a,Rako03}. They arise as a consequence of short-range
correlations or because of the occurrence of shocks. They have
a microscopic width which is finite on lattice scale and hence leads
to a boundary discontinuity when under hydrodynamic scaling
the lattice spacing is sent to zero. This is contrast to the boundary
layers observed here which remains finite even for vanishing lattice
spacing. Since there are no shocks in purely boundary-driven 
system, a possible explanation for the origin of the boundary
layers could be the presence of long-range
correlations in the stationary distribution.

Acknowledgement: Financial support by the Deutsche Forschungsgemeinschaft 
within the priority programme SPP1155 is 
gratefully acknowledged. We also thank Rosemary Harris, Dragi Karevski,
J\"org K\"arger 
and Henk van Beijeren for useful 
discussions. G.M.S thanks the Isaac Newton Institute for Mathematical Sciences
(Cambridge), where part of this work was done, for kind hospitality.

\newpage

\appendix
\section{Exact two-species self-diffusion coefficients on a finite lattice}
\label{selfdiff}

The self-diffusion coefficient is defined to be proportional to the asymptotic variance of the
displacement $x(t)$. More precisely, we define
\begin{align}
\label{dsdef}
D_s=\frac{1}{2}\lim_{t \to \infty} \frac{d}{dt}\left( \lav x^2 \rav - \lav x \rav^2 \right).
\end{align}
of a tagged particle in infinite space. For an equilibrium system not driven by external
forces one has $\lav x \rav=0$ and the variance reduces to the mean square displacement $\lav x^2 \rav$.
For finite time one cannot generally expect $D_s$ to be constant 
(anomalous diffusion), in particular for single-file 
systems with one species of particles $D_s$ vanishes asymptotically as $t^{-\frac{1}{2}}$, see \cite{Levi73,vanB83}, and for a mathematically
rigorous proof for the SEP \cite{Arra83}.

For a finite system defined on a ring with periodic boundary conditions the definition
\eqref{dsdef} becomes meaningless. For a lattice model we consider instead the number
of jumps $x^+$ in ``positive'' direction (say, clockwise) and the number
of jumps $x^-$ in the opposite direction. As displacement we then define the
quantity $x = x^+-x^-$. The diffusion coefficient may then be defined
by \eqref{dsdef}. In a periodic single-file system with $L$ sites $D_s$ is 
finite even for $t \to \infty$ and is proportional
to $1/L$ \cite{vanB83}. 
In the two-component symmetric simple exclusion process we one at first
sight sight expect two different
self-diffusion coefficients $D_{sA}$ and $D_{sB}$ for the $A$ and $B$ 
particles, respectively. In this section we show that
$D_{sA}=D_{sB}\equiv D_s$ by an exact calculation for $t \to \infty$ for
a finite system with periodic boundary conditions.
The calculation is done via a mapping of the 
exclusion process to the zero range process (ZRP) as follows.

\begin{figure}
\includegraphics[width=14cm]{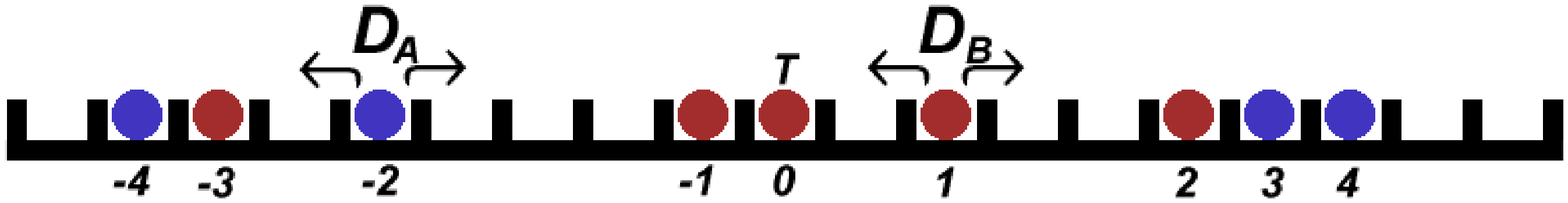}
\includegraphics[width=10cm]{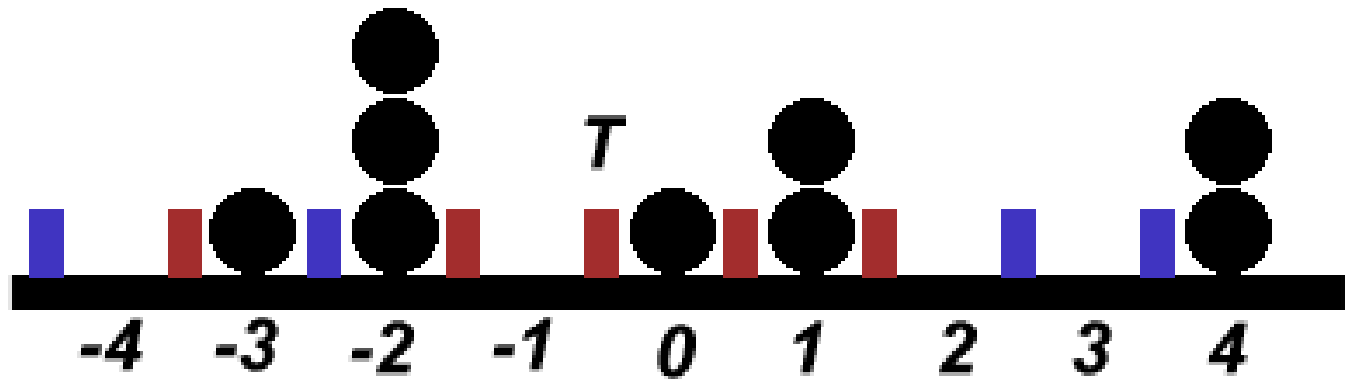}
\caption{Mapping of the SEP with two species of particles to the ZRP.
Particles in the SEP correspond to sites in the ZRP and interparticle
distances correspond to occupation numbers. The particle hopping
rate in the SEP maps to a hopping rate across bonds as indicated.}
\label{latticeThreeStateMapping}
\end{figure}

In the lattice gas picture the state of the system with length $L$ is 
characterized by a set of $L$ occupation numbers $y=(y_1,...,y_L)$ each of 
which can take values $-1$ (B-particle), $1$ (A-particle) or $0$ (vancancy). 
One defines a different lattice by labelling the $N$ particles as illustrated 
in Fig.~\ref{latticeThreeStateMapping}. The individual labels correspond to the site number of the
new lattice which, therefore, is of length $N$. Let $s=(s_1,...,s_N)$ be the configuration 
of the new lattice. Then $s_i=0,1,2,..$, counting the number of particles at site $i$
is defined as the number of 
empty sites to the right of particle $i$ in the SEP. This yields a particle system
without exclusion, named  zero range process (ZRP) \cite{Spit70}.
A jump of particle $i$ in the SEP to the right
induces a jump of a particle located on site $i$ of the ZRP to 
site $i-1$. Similarly, a jump to the left in the SEP corresponds to a jump
from $i-1$ to $i$ in the ZRP. The different hopping rates $D_A$, $D_B$ assigned
to a particle in the SEP are translated into different hopping 
rates across bonds in the ZRP. The symmetry in hopping of the SEP is translated into 
a symmetric hopping across a bond. 

In order to proceed with the calculation of the $A$-particle self-diffusion coefficient we
tag an $A$-particle and without loss of generality label it by zero. 
Let us now track the displacement of this particle set $x(0)=0$. Then the displacement $x(t)$ for the
particle located on the ring is the number of jumps to the left 
minus the number of jumps to the right, performed within the time span 0 and t. 
In terms of the ZRP this number corresponds to the total particle current 
across bond $(0,1)$. The location of a tagged particle is thus given by 
counting the number of jumps across a fixed bond in the ZRP. 

In order to set 
up the generator for the ZRP we define the single-site basis vectors for site 
$i$ as follows: 
An infinite row vector with a '$1$' at the first entry (zero elsewhere) is 
assigned to having no particle at 
site $i$. Provided site $i$ carries one particle, then a vector with '$1$' at 
the second entry (zero elsewhere) is assigned and so on. The basis vectors for 
the entire chain are as usual composed
by a tensor product of the single-site vectors. $A$ hopping event from site $i$
to $i+1$ is described by the combined action of the creation and annihilation
matrix $a^+_{i+1}$ and $a^-_{i}$. See appendix $B$ for a representation of the 
matrices and
details of the calculation. To each bond $(i-1,i)$ we need to assign a hopping rate $D_i=(D_A,D_B)$
which corresponds to the hopping rate of the $i$'th particle in the SEP.
This yields the following expression for the Hamiltonian:
\begin{align}
\label{hzrp}
 H_{ZRP}=-\sum_i D_i\left( a_i^- a_{i+1}^+ - d_i + a_i^+ a_{i+1}^- - d_{i+1} \right)
\end{align}
The diagonal part of the Hamiltonian which ensures conservation of probability is
constructed by the operators $d_i$. We remind the reader that the ground state
of this Hamiltonian gives the stationary probability distribution which is unique
for given total number of particles in the zero-range process.

For purposes that will immediately become clear we also introduce a non-equilibrium
ZRP where particle hopping across bond $(0,1)$ is asymmetric, i.e., a particle
hops from site 0 to site 1 with a rate $D_A q$ and with rate $D_A q^{-1}$ it
hops from site 1 to site 0. Notice that due to periodic boundary conditions
site 0 is identical with site $N$. We set $q = {\rm e}^{E/2}$ where $E$ plays
the role of a local driving force (measured in units of the thermal factor $k_BT$).
The generator of this driven process may be written
\begin{align}
\label{hzrpq}
 H_{ZRP}(E)= H_{ZRP} + V(E)
\end{align}
where
\begin{align}
\label{ve}
 V(E) = V=-D_A(q-1)(a_{0}^- a_1^+-d_0)-D_A(q^{-1} -1)(a_{0}^+ a_1^--d_1).
\end{align}
In this system a stationary current $j(E)$ will emerge. In the original
SEP this modification corresponds to a force acting on the tagged particle. 
The Einstein relation \cite{Katz84} then asserts that
\begin{align}
\label{einstein}
D_s = j'(0)
\end{align}
where the prime denotes the derivative w.r.t. the force $E$.

The proof of this relation \cite{Katz84} does not seem to be generally known 
and we outline the main ideas here. In order to track the number of jumps 
over bond $(-1,0)$ we follow the strategy of \cite{Gabi2004} and introduce the 
counting operators $X^+$ and $X^-$, counting clockwise and anticlockwise jumps, 
see also \cite{Harr07} for a more general description of counting processes. 
The counting operators act locally on an additional subspace with the action
\begin{align}
X_-|k>&=|k-1>\\
X_+|k>&=|k+1>.
\end{align}
Tracking the position of the tagged particle in the SEP is now implemented by 
the operator $X$ with eigenvalue $k$: $X|k>=k|k>$. The full Hamiltonian 
$H=H_{ZRP}+V$ may be split into a term $H_{ZRP}$, and a ``perturbation'' 
$V$ with $V=-D_A(X^+-1)a_{0}^- a_1^+-D_A(x^- -1)a_{0}^+ a_1^-$.

Due to the left-right symmetry of the hopping, $\lav x(t) \rav=0$. It remains 
to calculate the second moment of the displacement $x$ in the late-time limit
\begin{align}
\label{matrixelement}
\lav x^2 \rav=\lav s|X^2 e^{-Ht}|P^* \rav .
\end{align}
where $| P^* \rangle$ is the stationary probability vector.
This computation starts by expanding the exponential $\exp(-Ht)$ into a time-ordered 
Dyson series with respect to the perturbation $V$
\begin{align}
 e^{-(H_{ZRP}+V)t}=e^{-H_{ZRP}t}\left[1-
 \int_0^t d\tau_1 V(\tau_1) + \int_0^t d\tau_1 \int_0^{\tau_1} d\tau_2 V(\tau_1) V(\tau_2)-...\right]
\end{align}
with $V(\tau)=e^{H_{ZRP}\tau}Ve^{-H_{ZRP}\tau}$. 
When calculating the $n$'th 
moment of $x$ any terms of order than higher $n$ 
vanish identically in the matrix element \eqref{matrixelement}
which follows from the commutation relation $[X,X^\pm]=\pm X^\pm$ and
$\langle s | X^\pm=\langle s |$.
\begin{multline}
 <x^2>=-<s|X^2e^{-H_{ZRP}t} \int_0^t d\tau_1 V(\tau_1)|P^*> 
    \\+<s|x^2e^{-H_{ZRP}t} \int_0^t d\tau_1 \int_0^{\tau_1} d\tau_2 V(\tau_1) V(\tau_2) |P^*>
\end{multline}
and one finds
\begin{align}
\label{dsint}
D_{s}=D_A \lim_{t \to \infty}\left(
    \frac{D_A}{2}\left( \lav d_0 \rav + \lav d_1 \rav \right) -
    D_A\int_0^\infty dt \lav s|(d_0-d_1)e^{-Ht}(d_0-d_1)|P^* \rav \right).
\end{align} 
The proof of the Einstein relation \eqref{einstein} is completed by noting
that the same expression involving the time-integrated current-correlation 
function appears in the computation of $j'(0)$ using standard first-order
time-dependent perturbation theory from quantum mechanics.

For the actual computation of $D_s$ we therefore need $j(E)$ which
is computed in the next appendix.
From \eqref{solution} we read off by taking the derivative at $E=0$
\begin{align}
D_{s} = z_0 \left[ \sum_{i=1}{M-1} {D_i^{-1}} \right]^{-1}
\end{align} 
Using $M=N_A+N_B$ and  \eqref{fugacityrelation} this yields 
in terms of the parameters of the two-component SEP
\begin{align}
 \sum_{i=1}{M-1} {D_i^{-1}} = \frac{1}{L}
\left(\frac{\rho_A}{D_A} + \frac{\rho_B}{D_B}\right)
\end{align} 
from which \eqref{ds} follows.

\section{Quantum Hamiltonian formalism for the zero-range process}

The state of a single site $i$ is characterized by the occupation
number $s_i$. Let us first confine to the case of non-conserved 
particle number. For the infinite number of states one chooses the vector
representation with the single-site basis vectors
\begin{align}
\label{basis2}
|0>=
\left(
  \begin{array}{l}
    1 \\
    0 \\
    0 \\
    \vdots 
  \end{array}
\right), \quad
|1>=
\left(
  \begin{array}{l}
    0 \\
    1 \\
    0 \\
    \vdots
  \end{array}
\right), \quad
|2>
\left(
  \begin{array}{l}
    0 \\
    0 \\
    1 \\
    \vdots
  \end{array}
\right).
\end{align}
The basis vectors representing the state of the entire lattice with
$M$ is given by
the tensor product of the single-site vectors
\begin{align}
  |\eta>=|s_0>\otimes|s_1>\otimes...\otimes|s_{M-1}>.
\end{align}

In this basis the operation of creating ($a_i^+$) and deleting
a particle ($a_i^-$) at site $i$ is represented by the matrices
\begin{align}
a_i^+=\left(
  \begin{array}{lllll}
    0\quad  &       &       &       &\\
    1\quad  &\ddots &       &       &\\ 
            &\ddots &\ddots &       &\\
            &       &1      &0\quad &\\
            &       &       &\ddots &\ddots
  \end{array}
\right)
\quad
a_i^-=\left(
  \begin{array}{lllll}
    0\quad  &1\quad      &       &       &\\
            &\ddots      &\ddots &       &\\ 
            &            &\ddots &1\quad      &\\
            &            &       &0\quad      &\ddots\\
            &            &       &        &\ddots
  \end{array}
\right)
\end{align}
The operator $a_i^+ a_{i+1}^-$, hence, describes hopping from site $i+1$ to $i$. According
to the rules described in appendix A, conservation of probability demands to introduce
\begin{align}
d_i=\left(
  \begin{array}{lll}
    0\quad  &  & \\
       &1\quad & \\ 
       &  &\ddots \\
  \end{array}
\right).
\end{align}

The stationary state for $H_{ZRP}$ \eqref{hzrp} can be obtained by a product 
ansatz of the form
\begin{align}
|P^*>=
(1-z)
\left(
  \begin{array}{l}
    1 \\
    z \\
    z^2 \\
    \vdots
  \end{array}
\right)^{\otimes M}
\end{align}
where the fugacity $z$ is related to the ZR particle density via $z=\frac{c}{1+c}$.
For space-dependent hopping rates the same ansatz works with space-dependent
fagacity $z_i$. This fugacity is determined
by the stationarity condition
\begin{align}
H |P^*>= 0
\end{align}
which together with
\begin{align}
\label{recursion}
\lav d_i \rav = z_i
\end{align}
yields the recursion relation
\begin{align}
j = D_i (z_i - z_{i+1}) \quad i\neq 0\\
= D_0 (q z_0 - q^{-1} z_{1}).
\end{align}
Here we have assumed periodic boundary conditions with site $M \equiv 0$.

This recursion is solved in terms of the free parameter
$z_0$ by the relation
\begin{align}
\label{solution}
j = z_0 \frac{q  - q^{-1}}{D_0^{-1} + q^{-1} \sum_{i=1}^{M-1} D_i^{-1}}.
\end{align}
The fugacity $z_0$ fixes the local density at site $i=0$ and 
via \eqref{recursion} all other local densities. We remark
that $z_k - z_1 = j  \sum_{i=1}{k-1} {D_i^{-1}}$ which by the law
of large numbers implies a linear fugacity profile on macroscopic 
scale. For $q=1$ (vanishing driving field $E=0$) we have $j=0$ and
therefore constant fugacities $
z_k=z=c/(1+c)$.
In terms of the particle density in the underlying SEP with $N=M$
particles this gives
\begin{align}
\label{fugacityrelation}
z=1-\rho.
\end{align}


\begin{thebibliography}{10}

\bibitem{Karg92}
K\"arger J and Ruthven D M 1992 
\newblock {\it Diffusion in zeolites}
\newblock (John Wiley: New York)

\bibitem{Kukl96} 
Kukla V, Kornatowski J, Demuth D, Girnus I,
Pfeifer H, Rees L V C, Schunk S, Unger K and K\"arger J 1996
\newblock NMR studies of single-file diffusion in unidimensional channel zeolites
\newblock {\it Science} {\bf 272} 702

\bibitem{Keil00}
F. Keil, R. Krishna, M.-O. Coppens 2000
\newblock Modeling of diffusion in zeolites
\newblock {\it Rev. Chem. Eng.} {\bf 16} 71

\bibitem{Wei00} Wei Q-H, Bechinger C and Leiderer P 2000
\newblock Single-File diffusion of colloids in one-dimensional channels
\newblock {\it Science} {\bf 287} 625

\bibitem{Perk94} Perkins T T, Smith D E and Chu S 1994
\newblock Direct observation of tube-like motion of a  single polymer-chain
\newblock {\it Science} {\bf 264} 819

\bibitem{Schu97} Sch\"utz G M 1997
\newblock The Heisenberg chain as a dynamical model
for protein synthesis - Some theoretical and experimental results
\newblock {\it  Int. J. Mod. Phys. B} {\bf 11} 197

\bibitem{Lipo05}  Lipowsky R and Klumpp S 2005
\newblock 'Life is motion': multiscale motility of molecular motors
\newblock  {\it  Physica A} {\bf 352}(1) 53

\bibitem{Nish05}
Nishinari K, Okada Y, Schadschneider A and Chowdhury D 2005
\newblock Intracellular transport of single-headed molecular motors KIF1A 
\newblock  {\it  Phys. Rev. Lett.} {\bf 95} 118101.

\bibitem{Levi73}
Levitt D G 1973
\newblock Dynamics of a Single-File Pore: Non-Fickian Behavior
\newblock  {\it  Phys. Rev. A} {\bf 8} 3050

\bibitem{vanB83} van Beijeren H, Kehr K W and Kutner R 1983 
\newblock Diffusion in concentrated lattice gases. III. Tracer diffusion on a
one-dimensional lattice
\newblock {\it Phys. Rev. B}  {\bf 28}  5711

\bibitem{Spoh83}
Spohn H 1983
\newblock Long-range correlations for stochastic lattice gases in a non-equilibrium
steady state
\newblock {\it J. Phys. A} {\bf 16} 4275

\bibitem{Derr02}
Derrida B, Lebowitz J L, and Speer E R 2002
\newblock Large deviation of the density profile in the steady state of the open 
symmetric simple exclusion process 
\newblock {\it J. Stat. Phys.} {\bf 107} 599

\bibitem{Bert02}
Bertini L, De Sole A, Gabrielli D, Jona-Lasinio G and Landim C 2002
\newblock Macroscopic fluctuation theory for stationary non-equilibrium states 
\newblock {\it J. Stat. Phys.} {\bf 107} 635

\bibitem{Ligg99}  
Liggett T M 1999 
\newblock {\it Stochastic Models of Interacting
Systems: Contact, Voter and Exclusion Processes} 
\newblock (Berlin: Springer)

\bibitem{Schu01}  
Sch\"{u}tz G M 2001 
\newblock Exactly solvable models for many-body systems far from equilibrium, 
in: {\it Phase Transitions and Critical Phenomena.} Vol. 19,
eds C Domb and J Lebowitz
\newblock (London: Academic Press)

\bibitem{Quas92}
Quastel J 1992
\newblock Diffusion of Color in the simple exclusion process
\newblock {\it Comm. Pure Appl. Math.}  {\bf 45} 623

\bibitem{Brza06}
Brzank A and Sch\"utz G M 2006
\newblock Boundary-induced bulk phase transition and violation of Fick's law in 
two-component single-file diffusion with open boundaries
\newblock {\it Diffusion Fundamentals} {\bf 4} 7.1-7.12

\bibitem{Czap02} Czaplewski K F, Reitz T L, Kim Y J and Snurr R Q 2002
\newblock One-dimensional zeolites as hydrocarbon traps 
\newblock {\it Micropor. Mesopor. Mater.} {\bf 56} 55

\bibitem{Rubi87} Rubinstein M 1987 
\newblock Discretized model of entangled-polymer dynamics
\newblock {\it Phys. Rev. Lett.} {\bf 59} 1946

\bibitem{Duke89} Duke T A J 1989 
\newblock Tube model of field-inversion electrophoresis
\newblock {\it Phys. Rev. Lett.} {\bf 62} 2877

\bibitem{Toth03} Toth B and Valko B 2003 
\newblock Onsager relations and Eulerian hydrodynamic limit for systems 
with several conservation laws 
\newblock {\it J. Stat. Phys.}  {\bf 112}(3-4) 497

\bibitem{Popk03b} Popkov V and Sch\"utz G M
\newblock Shocks and excitation dynamics in a driven diffusive two-channel system 
\newblock {\it J. Stat. Phys.}  {\bf 112}(3-4) 523

\bibitem{Schu03}
Sch\"utz G M 2003
\newblock Critical phenomena and universal dynamics in one-dimensional 
driven diffusive systems with two species of particles 
\newblock {\it J. Phys. A} {\bf 36} R339

\bibitem{Asla00}
Aslangul C 2000
\newblock Single-file diffusion with random diffusion constants 
\newblock {\it J. Phys. A} {\bf 33} 851

\bibitem{Arra83} 
Arratia R 1983
\newblock The Motion of a Tagged Particle in the Simple Symmetric 
Exclusion System in Z 
\newblock {\it Ann. Prob.} {\bf 11} 362

\bibitem{Vase06}
Vasenkov S, Sch\"uring A and Fritzsche S 2006
\newblock Single-file diffusion near channel boundaries 
\newblock {\it Langmuir} {\bf 22} 5728

\bibitem{Kaerg2002}
Vasenkov S and K\"arger J
\newblock Different time regimes of tracer exchange in single-file systems 
\newblock {\it Phys. Rev. E} {\bf 66} 052601

\bibitem{Leeuwen2003}
Drzewinski A, Carlon E and van Leeuwen J M J 2003
\newblock Pulling reptating polymers by one end: 
Magnetophoresis in the Rubinstein-Duke model 
\newblock {\it  Phys. Rev. E} {\bf  68} 061801

\bibitem{Anta00} 
Antal T and Sch\"utz G M 2000 
\newblock Asymmetric exclusion process with next-nearest-neighbor interaction: 
Some comments on traffic flow and a nonequilibrium reentrance transition 
\newblock {\it Phys. Rev. E} {\bf 62} 83

\bibitem{Popk03a} 
Popkov V, R\'akos A, Willmann R D, Kolomeisky A B and Sch\"utz G M 2003 
\newblock Localization of shocks in driven diffusive systems 
without particle number conservation 
\newblock {\it Phys. Rev. E} {\bf 67} 066117 Part 2 JUN 2003 

\bibitem{Rako03}    
Rakos A, Paessens M and Sch\"utz G M 2003 
\newblock Hysteresis in one-dimensional reaction-diffusion systems 
\newblock {\it Phys. Rev. Lett.} {\bf 91} 238302

\bibitem{Spit70} 
Spitzer F 1970 
\newblock Interaction of Markov Processes
\newblock {\it Adv. Math.} {\bf 5} 246

\bibitem{Gabi2004}
Sch\"onherr G and Sch\"utz G M 2004
\newblock Exclusion process for particles of arbitrary extension: 
hydrodynamic limit and algebraic properties
\newblock {\it J. Phys. A} {\bf 37} 8215

\bibitem{Harr07}
Harris R J and Sch\"utz G M 2007
\newblock Fluctuation theorems for stochastic dynamics
\newblock cond-mat/0702553, to appear in JSTAT

\bibitem{Katz84} 
Katz S, Lebowitz J L and Spohn H 1984 
\newblock Nonequilibrium steady states of stochastic lattice gas models 
of fast ionic conductors
\newblock {\it J. Stat. Phys.} {\bf 34} 497

\end{thebibliography}
\end{document}